\documentclass[11pt]{article}
\usepackage{hyperref}
\usepackage{amsmath}
\usepackage{amssymb}
\renewcommand{\title}[1]{%
    \bigskip%
    \begin{center}%
    \Large\bf #1%
    \end{center}%
    \vskip .2in}

\renewcommand{\author}[1]{%
    {\begin{center}
    #1
    \end{center}}}
\newcommand{\address}[1]{\vspace{-1.7em}\vspace{0pt}
    {\begin{center}
    \it #1
    \end{center}}}

\begin{document}


\title{Localisation of the Galilean symmetry and dynamical realisation of Newton-Cartan geometry}

\author
{
Rabin Banerjee  $\,^{\rm a,b}$,
Arpita Mitra    $\,^{\rm a, e}$,
Pradip Mukherjee $\,^{\rm c,d}$}
\address{$^{\rm a}$S. N. Bose National Centre 
for Basic Sciences, JD Block, Sector III, Salt Lake City, Kolkata -700 098, India }

\address{$^{\rm c}$Department of Physics, Barasat Government College,\\Barasat, West Bengal

 }

\address{$^{\rm b}$\tt rabin@bose.res.in}
\address{$^{\rm d}$\tt mukhpradip@gmail.com}
\address{$^{\rm e}$\tt arpita12t@bose.res.in}

\begin{abstract}
  Newtonian gravity was formulated as a geometrodynamic theory as far back in 1930s by Elie Cartan in what is named aptly as Newton Cartan space time. Though there are several approaches of realizing the algebraic structure of the Newton Cartan geometry from a contraction of the relativistic results, a dynamical (field theoretic) realization of it is lacking. In this paper we present such a realization from the localisation of the Galilean Symmetry of nonrelativistic matter field theories. 
\end{abstract}

\section{Introduction}
Newtonian gravity remains a celebrated theory of gravitation with a long list of excellent agreements between theory and experiment. However, it can only be an approximate theory because it is invariant under the galilean group and not under the Poincare group of transformations. Einstein's general theory of relativity (GTR) provided a beautiful theory of gravitation which is consistent with the special theory of relativity. It is a unique theory in the sense that spacetime became dynamic where gravity is equated with the curvature of spacetime. GTR is naturally framed on the four dimensional spacetime manifold. Since then the geometry of the spacetime manifold has been extensively studied. The Riemannian spacetime of GTR is generalised to Riemann - Cartan spacetime which is endowed with torsion in addition to curvature.

         An alternative approach to theories of gravitation was pioneered by  Utiyama, Kibble and Sciama \cite{Utiyama:1956sy, Kibble:1961ba, sciama, BH}. They utilised the gauge principle so successful in other branches of modern theoretical physics to the Poincare group. The resulting theory is called Poincare gauge theory (PGT). The theory is constructed by localising the Poincare transformations in Minkowski space. This theory provides a source of diverse ramifications in classical gravity \cite{Blagojevic:2002du}. It also enables one to formulate gravity theories in noncommutative spacetime \cite{Chamseddine:2000si, Calmet:2005qm, Mukherjee:2006nd, Banerjee:2007th}. The theories invariant under local Poincare transformations can be identified with diffeomorphism invariant theories in curved spacetime. The geometric interpretation of PGT, based on local Lorentz transformations (LLT) and general coordinate transformations or diffeomorphisms (diff) is buttressed by the careful demonstration of equivalence between different categories of transformation \cite{DRda}, \cite{BGMR} and is also shown to be consistent with gauging higher order matter theories \cite{M}. 

  Following the geometric setting of Einstein's theory attempts were made to formulate Newtonian gravity on a similar basis. Elie Cartan developed a four-dimensional geometrical formulation of Newton's gravitational theory in the twenties \cite{Cartan}. 
Trautman \cite{TrautA},  Dautcourt \cite{Daut, Daut1}, Ehlers\cite{EHL} and others approached the problem as a non-relativistic limit of  general relativity. Of course in Newtonian dynamics time flows universally in all inertial frames of reference (IFR). Naturally, the gravitational structure is defined in terms of an affine connection compatible with the temporal flow $t_{\mu}$ and a rank-three spatial metric $h^{\mu\nu}$. The connection of PGT with the Riemann - Cartan spacetime mentioned earlier raises the natural question -- is it possible to formulate the geometric structure of Newtonian gravity from localisation of galilean symmetry? In recent past several authors have approached the problem from different angles. A 3-dimensional gauging of the galilean symmetry a la Utiyama was done for a  non - relativistic point particle model \cite{PLP}. Also notable is the fact that in \cite{PLP} the dynamics of the gauge fields was obtained by a contraction of the corresponding relativistic particle action and not by direct gauge theoretic approach. Elsewhere \cite{ BP}, the problem of gauging the galilean group has been addressed from group theoretic approach. However, localisation of the symmetries of a galilean invariant field theory in line with the gauging of Poincare symmetries \cite{BH} and the subsequent realization of the Newton-Cartan geometry remains to be analyzed in detail. Recently we have discussed the localisation of the Galileo symmetric models in connection with nonrelativistic diffeomorphism in 3-space \cite{BMM}. One naturally wonders whether the localised theory can be interpreted as a geometric theory in 4-dimensional Newton-Cartan spacetime? That this is possible will be shown here.

In the present paper we would like to address the above problem following the localisation procedure introduced in  \cite{BMM}. Evidently a more elaborate treatment of this procedure will be essential here. In what follows we will use the analogy to the Utiyama method of the realisation of the Riemann-Cartan spacetime from Poincare Gauge Theory (PGT) in letter and spirit. This will naturally lead, step by step, to an explicit realisation of the Newton-Cartan space. 

   A thorough treatment of the localisation of the galilean symmetry sets up the stage to formulate the localised theory in 4-dimensional space time. The basic fields introduced during the localisation process are regrouped in an invertible $4 \times 4$ master matrix. The elements of this matrix and its inverse provide the necessary components to reconstruct  the Newton-Cartan structures. A four dimensional manifold ${\cal{M}}$ is defined where the connections between the local and the global bases will be given by vierbeins obtained from our pool of fields. At this point a geometrical interpretation is established which has of course been motivated by our experience of nonrelativistic diffeomorphism in space. These vierbeins are then used to generate the degenerate metrics of the Newton-Cartan space. Also a connection is derived from the vierbein postulate. These constructions have to go through several stringent tests. On one hand their transformation rules are determined by the localisation procedure outlined in \cite{BMM} and further elaborated here. On the other side, in the new geometric {\it{avatars}}, they have to satisfy definite tensorial transformation rules. On top of it they have to satisfy the salient geometric properties of the Newton-Cartan spacetime. Successful verification of all these properties sets our construction on a firm footing. We thus have a dynamical realization of the Newton-Cartan spacetime without taking any help from the relativistic theories -- be that the contraction of the relativistic particle dynamics as in \cite{PLP} or, the contraction of the Poincare algebra as in \cite{BP}. 

 In section 2 the localisation of galilean invariance in a general nonrelativistic matter theory will be discussed. This is an elaboration of the results provided in \cite{BMM} and contains new details of calculations which highlight the geometrical perspective more clearly. Section 3 contains the main results of this paper where, guided by geometric analogy, we construct the basic structures of the Newton-Cartan spacetime.
The final section as usual contains the concluding remarks.
\section{Localisation of Galilean Symmetry}
This section contains a very brief summary of our earlier work \cite{BMM} that will be needed in this paper. More details can be found in \cite {BMM}.

Localizing the Galilean transformation as,
\begin{equation}
\xi^{0}=-\epsilon\left(t\right),~~~~~~\xi^i = \eta^i\left(t, {\bf{r}}\right) - v^i\left(t, {\bf{r}}\right)t
\label{localgalilean}
\end{equation}
we convert a globally Galilean symmetric model,
\begin{equation}
S = \int dt d^3 x {\cal{L}}\left(\phi,\partial_t{\phi}, \partial_k{\phi}\right)
\label{action}
\end{equation}
to,
\begin{equation}
S = \int dt d^3x \frac{M}{\theta}{\cal{L}}\left(\phi, \nabla_t\phi, \nabla_a\phi\right)
\label{localaction}
\end{equation}
which is invariant under (\ref{localgalilean}). M is given by,
\begin{equation}
 M = det{\Lambda_k}^a
 \label{M}.
\end{equation}
where  ${\Lambda_k}^a$ is the inverse of ${\Sigma_a}^{k}$ which enters in the definition of the covariant derivatives $\nabla_t, \nabla_a$ as,
\begin{equation}
\nabla_t\phi=\theta(D_t \phi+\Psi^k D_k\phi)
\label{finalcov}
\end{equation}
\begin{equation}
\nabla_a\phi={\Sigma_a}^{k}D_k\phi.
\label{nab}
\end{equation}
where,
\begin{eqnarray}
D_k\phi=\partial_k\phi+iA_k\phi\nonumber\\
D_t\phi=\partial_t\phi+iA_t\phi \label{firstcov}
\end{eqnarray}
To achieve local galilean invariance new fields $A_t(t, {\bf{r}}), A_k(t, {\bf{r}}), \theta (t)$, $\Psi^k(t, {\bf{r}}), {\Sigma_a}^k(t, {\bf{r}})$ and ${\Lambda_k}^a(t, {\bf{r}})$ are introduced whose required transformations are given by,
\begin{align}
\delta_0A_t &= m{\dot{v}}^ix_i+\epsilon{\dot{A}}_t+\dot{\epsilon}A_t-\eta^i
\partial_iA_t+v^i t\partial_i A_t-{\dot{\eta}}^iA_i+{\dot{v}}_i tA_i+v^i A_i +m{\Lambda_k}^a\Psi^k v_a
\nonumber\\
{\delta}_0 A_{k} &= -{\partial}_k {\eta}^i A_i +t{\partial}_k v^iA_i + \epsilon \dot{A}_k - {\eta}^i {\partial}_iA_k +tv^i{\partial}_iA_k + mv_k +  m{\partial}_kv^ix_i-m{\Lambda_k}^a v_a\nonumber\\
\delta_0\theta &=-\theta\dot{\epsilon}+\epsilon\dot{\theta}\nonumber\\
\delta_0\Psi^k &=\epsilon{\dot{\Psi}}^k+\dot{\epsilon}\Psi^k-\frac{1}{\theta}v^b{\Sigma}^{k}_{b}+\frac{\partial}{\partial t}({\eta}^k-tv^k)-(\eta^i-v^i t) \partial_i\Psi^k-t\Psi^i\partial_i v^k+\Psi^i\partial_i\eta^k\nonumber\\
\delta_0 {\Sigma_a}^{k} &=\epsilon{\dot{\Sigma}_a}^{k}+ {\Sigma_a}^{i}\partial_{i}\eta^{k}-t{\Sigma_a}^{i}\partial_{i}v^{k} -
{\omega_a}^b{\Sigma_b}^{k} - \eta^{i}\partial_{i}{\Sigma_a}^{k}+t v^{i} \partial_{i} {\Sigma_a}^{k}\nonumber\\\delta_0 {\Lambda_k}^a &=\epsilon{\dot{\Lambda}}_k{}^{a}- {\Lambda_l}^{a}\partial_{k}\eta^{l}+t{\Lambda_l}^{a}\partial_{k}v^{l}-
{\omega^a}_{c}{\Lambda_k}^{c} - \eta^{i}\partial_{i}{\Lambda_k}^{a}+t v^{i} \partial_{i} {\Lambda_k}^{a}
\label{delth}
\end{align}
\section{Dynamical Construction of Newton-Cartan Geometry} 
In the previous section we have seen that our method of localisation of the galilean symmetry leads to models which are 3-dimensional diffeomorphism invariant. The fields ${\Lambda_k}^a$ act as vierbein in space through which we have defined the spatial metric. Note that here we were confined to spatial transformations only. However the geometric interpretation of our galilean gauge theory is far reaching. Just as Poincare gauge theory leads to the Riemann-Cartan spacetime \cite{BH, Blagojevic:2002du} the galilean gauge theory will be shown to reproduce the Newton-Cartan spacetime. So far the gauge procedure employed to realize the same depended on the contraction of Poincare algebra \cite{BP} and was devoid of any underlying dynamical structure. In contrast our method does not presuppose the relativistic spacetime. 

\subsection{Structure of Newton-Cartan Spacetime }
Before beginning the actual construction it will be useful to review the salient features of the Newton-Cartan spacetime \cite{Daut, Daut1, BP, DH}. It is a four dimensional manifold ${\cal{M}}$ endowed with two degenerate metrics of rank 3 and rank 1 respectively. It is convenient to take a degenerate spatial metric $h^{\rho\sigma}$ of rank 3 and a degenerate temporal vierbein $\tau_{\mu}$ of rank 1. A connection $\Gamma^{\mu}_{\nu\lambda}$ is assumed with respect to which the following metricity conditions hold,
\begin{align}
\nabla_\mu h^{\rho\sigma}&=\partial_{\mu}h^{\rho\sigma}+
\Gamma_{\nu\mu}^{\rho}h^{\nu\sigma}+\Gamma_{\nu\mu}^{\sigma}h^{\nu\rho}=
 0\notag\\\nabla_\mu \tau_{\nu}&= \partial_\mu \tau_{\nu}-\Gamma_{\mu\nu}^{\sigma}\tau_\sigma= 0.
\label{metricity}
\end{align}
To get an explicit form of the connection $\Gamma^{\mu}_{\nu\lambda}$ we also require to introduce the covariant quantities $h_{\mu\nu}$ and $\tau_\mu$. These are defined by the following properties
\begin{align}
 h^{\mu\nu}h_{\nu\rho} &= \delta^{\mu}_{\rho}-\tau^{\mu}\tau_{\rho},~~~~\tau^{\mu}\tau_{\mu} = 1,\nonumber\\
 ~~~h^{\mu\nu}\tau_{\nu} &= 0,~~~~ h_{\mu\nu}\tau^{\nu} = 0.
\label{inverse}
\end{align}
Using these the connection can be written as,
\begin{align}
{\Gamma^\sigma}_{\mu\nu} & = \tau^{\sigma}\partial_{(\mu}\tau_{\nu)} +
\frac{1}{2}h^{\sigma\rho} \Bigl(\partial_{\nu}h_{\rho\mu} + \partial_{\mu}h_{\rho\nu} - \partial_{\rho}h_{\mu\nu}\Bigr)
\label{ncm}
\end{align}
But the connection $\Gamma_{\mu\nu}^{\rho}$ is not uniquely determined by the metric compatibility conditions (\ref{metricity}). They are also preserved under the shift,
\begin{equation}
\Gamma^{\rho}_{\mu\nu} \rightarrow \Gamma^{\rho}_{\mu\nu} + h^{\rho\lambda}K_{\lambda(\mu}\tau_{\nu)}
\label{connectionshift}
\end{equation}
where $K_{\mu\nu}$ is an arbitrary two-form \cite{Daut}. Now it is possible to write the most general form of the metric compatible symmetric connection using this arbitrary two-form and (\ref{ncm}) as \cite{Daut, BP},
\begin{align}
{\Gamma^\sigma}_{\mu\nu} & = \tau^{\sigma}\partial_{(\mu}\tau_{\nu)} +
\frac{1}{2}h^{\sigma\rho} \Bigl(\partial_{\nu}h_{\rho\mu} + \partial_{\mu}h_{\rho\nu} - \partial_{\rho}h_{\mu\nu}\Bigr)+ h^{\sigma\lambda}K_{\lambda(\mu}\tau_{\nu)}\,.
\label{covariantconnection}
\end{align}
\subsection{Geometrical Interpretation of The Galilean Gauge Theory -- Consruction of The Newton-Cartan Geometry}
We now proceed to discuss the realization of the Newton-Cartan geometry from our method of localisation of the galilean symmetry. In the present section we will consider the geometric interpretation of the localisation of galilean symmetry in its full glory. We define a four dimensional manifold with `coordinates' $x_0, x_1, x_2, x_3$ and set up local and global frames. Now redefining the new fields introduced in the previous section as,
\begin{equation}
\theta = {\Sigma_0}^0,~~\theta\Psi^k ={\Sigma_0}^k.
\label{tts}
\end{equation}
and putting ${\Sigma_a}^0=0$ we can construct the $4\times 4$ invertible matrix,
\begin{equation}
{\Sigma_\alpha}^{\mu}=\begin{pmatrix}
\theta & \theta\Psi^k \\
0 & {\Sigma_a}^k
\end{pmatrix}
\label{Smatrix1}
\end{equation}
where ${\Sigma_a}^k$ has already been introduced in (\ref{nab}). The inverse matrix ${\Lambda_\mu}^\alpha$ satisfies,
\begin{equation}
{\Sigma_\alpha}^{\mu}\Lambda_{\mu}{}^{\beta}=\delta^{\beta}_{\alpha},~~~
{\Sigma_\alpha}^{\mu}\Lambda_{\nu}{}^{\alpha}=\delta^{\mu}_{\nu}
\label{sila}
\end{equation}
The spatial part $\Lambda_k{}^a$ is the inverse of $\Sigma_a{}^k$ as may be stated in previous section. Note that we are denoting the local coordinates by the initial Greek letters i.e. $\alpha,\beta $ etc. whereas the global coordinates are denoted by letters from the middle of the Greek alphabet i.e.  $\mu,\nu $ etc. From (\ref{tts}) and (\ref{delth}) the transformation of ${\Sigma_0}^k$ follows,
\begin{equation}
\delta_0 {\Sigma_0}^{k}=\epsilon{\dot{\Sigma}_0}^{k}+ {\Sigma_0}^{i}\partial_{i}(\eta^{k}-tv^{k}) - \eta^{i}\partial_{i}{\Sigma_0}^{k}+t v^{i} \partial_{i} {\Sigma_0}^{k}
+\partial_0\left(\eta^k - tv^k\right)\theta + v^b{\Sigma_b}^{k}
\label{delth2}
\end{equation}
Using the definition of $\xi^i$ from (\ref{localgalilean}) the transformations (\ref{delth}) can be simplified to,
\begin{align}
\delta_0 {\Sigma_0}^{k} &= -\xi^\nu {\partial_\nu{\Sigma}_0}^{k}+ {\Sigma_0}^{\nu}\partial_{\nu}\xi^{k} - v^b{\Sigma_b}^{k}\nonumber\\
\delta_0 {\Sigma_a}^{k} &= -\xi^\nu {\partial_\nu{\Sigma}_a}^{k}+ {\Sigma_a}^{\nu}\partial_{\nu}\xi^{k} -\omega_a{}^b{\Sigma_b}^{k}
\label{delth3}
\end{align}
 Similarly we can work out,
\begin{align}
\delta_0 {\Lambda_0}^{a} &= -\xi^\nu {\partial_\nu{\Lambda}_0}^{a}- {\Lambda_\nu}^{a}\partial_{0}\xi^{\nu} + v^a{\Lambda_0}^{0}\nonumber\\
\delta_0 {\Lambda_k}^{a} &= -\xi^\nu {\partial_\nu{\Lambda}_a}^{k}- {\Lambda_\nu}^a\partial_{k}\xi^{\nu}+{\omega_c}^a{\Lambda_k}^{c}
\label{delth4}
\end{align}
Our next task is to show that the 4-dimensional spacetime manifold endowed with the matrix $\Sigma_{\alpha}{}^{\mu}$ and its inverse $\Lambda_{\mu}{}^{\beta}$ has the features of the Newton-Cartan geometry.
 With this point in view we write down a degenerate spatial metric $h^{\mu\nu}$ of rank 3 and a degenerate temporal vierbein $\tau_{\mu}$ of rank 1 in the following way
 \begin{equation}
h^{\mu\nu}={\Sigma_a}^{\mu}{\Sigma_a}^{\nu}
\label{spm}
\end{equation}
and
\begin{equation}
\tau_{\mu}={\Lambda_\mu}^{0}~~~({\Lambda_k}^{0}=0, {\Lambda_0}^{0}\neq 0)
\label{tem}
\end{equation}
From the form variations of the basic fields (see (\ref{delth})) we get,
\begin{equation}
\delta_0 h^{\mu\nu} =-\xi^{\rho} \partial_{\rho} h^{\mu\nu}+h^{\rho\nu}\partial_{\rho}\xi^{\mu}+h^{\rho\sigma}\partial_{\sigma}\xi^{\mu}
\label{diff} 
\end{equation}
Similar results can be obtained for $\delta_0 \tau_{\mu}$. 
\begin{equation}
\delta_0 \tau_{\mu}=\delta_0 \Lambda_{\mu}{}^0=\delta_0 \Lambda_0{}^0=-\Lambda_0{}^0\partial_0 \xi^0+\xi^0\partial_0 \Lambda_0{}^0
\end{equation}
Using these relations it is easy to show that they have correct tensorial properties,
\begin{equation}
 h^{\mu\nu}\left(x^{\prime}\right) = \frac{\partial x^{\prime\mu}}{\partial x^{\rho}}\frac{\partial x^{\prime\nu}}{\partial x^{\sigma}}h^{\rho\sigma}(x)
\label{diffh} 
\end{equation}
and
\begin{equation}
 \tau_{\mu}\left(x^{\prime}\right) = \frac{\partial x^{\rho}}{\partial x^{\prime\mu}}\tau_{\rho}(x)
\label{difft} 
\end{equation}
The quantities ${\Sigma_\alpha}^{\mu}$ and $\Lambda_{\mu}{}^{\alpha}$ can be considered as direct and inverse vierbein. The affine connection $\Gamma_{\nu\mu}^{\rho}$ and the spin connection $A_\mu^{\alpha\beta}$ are introduced through the vierbein postulate \cite{BP}, 
\begin{equation}
\nabla_\mu{\Lambda_\nu}^{\alpha} = \partial_{\mu}{\Lambda_\nu}^{\alpha} - \Gamma_{\nu\mu}^{\rho}{\Lambda_\rho}^{\alpha}
+A^{\alpha}{}_{\mu\beta}{\Lambda_\nu}^{\beta} =0\,. 
 \label{P}
\end{equation}
For $\alpha=0$ we find,
\begin{equation}
\nabla_{\mu} \Lambda_{\nu}{}^0= \partial_{\mu}\Lambda_{\nu}{}^0 - \Gamma_{\nu\mu}^{\rho}\Lambda_{\rho}{}^0+A^{0}{}_{\mu\beta}\Lambda_{\nu}{}^{\beta} =0\,. 
 \label{P1}
\end{equation}
which implies,
\begin{equation}
 \partial_{\mu}\Lambda_{\nu}{}^0 - \Gamma_{\nu\mu}^{\rho}\Lambda_{\rho}{}^0=0
\end{equation}
As $A_{\mu}{}^{0\beta}$ vanishes for galilean transformation we reproduce the metricity condition (\ref{metricity}) for $\tau_\mu$ using (\ref{tem}).

 The proof of the metricity condition for $h^{\mu\nu}$ is a little bit involved. From(\ref{sila}) and (\ref{P}) it can be shown that,
\begin{equation}
\partial_{\mu}\Sigma_{\delta}{}^\sigma-A_{\mu}{}^{\beta}{}_{\delta}\Sigma_{\beta}{}^\sigma =-\Gamma_{\nu\mu}^{\sigma}\Sigma_{\delta}{}^\nu
\label{P2}
\end{equation}
Considering $\delta=a, \beta=b $ we get,
\begin{equation}
\partial_{\mu}\Sigma_{a}{}^\sigma-A_{\mu}{}^{b}{}_{a}\Sigma_{b}{}^\sigma =-\Gamma_{\nu\mu}^{\sigma}\Sigma_{a}{}^\nu
\label{Ploc}
\end{equation}
Multiplying $\Sigma_{a}{}^{\rho}$ to (\ref{Ploc}) gives,
\begin{equation}
\Sigma_{a}{}^{\rho}\partial_{\mu}\Sigma_{a}{}^\sigma-A_{\mu}{}^{b}{}_{a}\Sigma_{a}{}^{\rho}\Sigma_{b}{}^\sigma =-\Gamma_{\nu\mu}^{\sigma}\Sigma_{a}{}^{\rho}\Sigma_{a}{}^\nu
\label{Plocmu}
\end{equation}
Then we interchange the indices $\rho, \sigma$,
\begin{equation}
\Sigma_{a}{}^{\sigma}\partial_{\mu}\Sigma_{a}{}^\rho-A_{\mu}{}^{b}{}_{a}\Sigma_{a}{}^{\sigma}\Sigma_{b}{}^\rho =-\Gamma_{\nu\mu}^{\rho}\Sigma_{a}{}^{\sigma}\Sigma_{a}{}^\nu
\label{Plocmu11}
\end{equation}
Adding (\ref{Plocmu}) with (\ref{Plocmu11}) and using the antisymmetric property of $A_{\mu}{}^{a b}$ leads to the first metricity condition of (\ref{metricity}). 

Thus we can conclude that our constructions of $h^{\mu\nu}$ (\ref{spm}), $\tau_\mu (\ref{tem})$ and $\Gamma_{\nu\mu}^{\rho} (\ref{P})$ satisfy the metric compatibility conditions (\ref{metricity}).

Our next task is to entirely express the connection in terms of the metric. For this we require the covariant metric $h_{\mu\nu}$ and contravariant ${\tau^\mu}$. Let
\begin{equation}
h_{\nu\rho}=\Lambda_{\nu}{}^{a} \Lambda_{\rho}{}^{a}
\label{spm2}
\end{equation}
and
\begin{equation}
\tau^{\rho}={\Sigma_0}^{\rho}.
\label{tm2}
\end{equation}
Using (\ref{spm}) and (\ref{tem}) we immediately get,
\begin{align}
h^{\mu\nu}\tau_\nu &={\Sigma_a}^{\mu}{\Sigma_a}^{\nu} \Lambda_{\nu}{}^{0}\notag\\
&={\Sigma_a}^{\mu}\delta^0_a\notag\\ &=0
\end{align}
Also the identifications (\ref{tm2})and (\ref{tem}) show that 
$$\tau^\mu\tau_\mu = 1.$$From the definitions (\ref{spm2}) and (\ref{tm2}) we find
\begin{align}
h_{\mu\nu}\tau^\nu &=  {\Lambda_{\mu}}^a {\Lambda_\nu}^a {\Sigma_0}^\nu\notag\\ &=  {\Lambda_{\mu}}^a\delta_0^a\notag\\ &=0
\end{align}
Finally we can verify that $$h^{\mu\lambda}h_{\lambda\nu} = \delta^\mu_\nu - \tau^\mu\tau_\nu.$$
This completes the realization of the Newton-Cartan algebra.

Finally, it will be shown that the connection $\Gamma_{\nu\mu}^{\rho}$ defined in (\ref{P}) can be put in the general form (\ref{covariantconnection}) as required in the Newton-Cartan construction. We can write from (\ref{P}), 
\begin{equation}
\Gamma_{\nu\mu}^{\rho} = \partial_{\mu}{\Lambda_{\nu}}^\alpha {\Sigma_\alpha}^{\rho}
+A^{\alpha}{}_{\mu\beta}{\Lambda_{\nu}}^\beta{\Sigma_\alpha}^{\rho}\label{vpcon}
\end{equation}
Assuming that the connection is symmetric (\ref{vpcon}) can be written as,
\begin{align}
\Gamma_{\nu\mu}^{\rho} &=\frac{1}{2}[\Gamma_{\nu\mu}^{\rho}+\Gamma_{\mu\nu}^{\rho}]\notag\\
 &=\frac{1}{2}[\partial_{\mu}{\Lambda_{\nu}}^{0} {\Sigma_0}^{\rho}+\partial_{\nu}{\Lambda_{\mu}}^0 {\Sigma_0}^{\rho}+\partial_{\mu}{\Lambda_{\nu}}^a {\Sigma_a}^{\rho}+\partial_{\nu}{\Lambda_{\mu}}^a {\Sigma_a}^{\rho}\notag\\ &+A^{a}{}_{\mu 0}{\Lambda_{\nu}}^{0}{\Sigma_\alpha}^{\rho}+A^{a}{}_{\nu 0}{\Lambda_{\mu}}^0{\Sigma_a}^{\rho}+A^{a}{}_{\mu b}{\Lambda_{\nu}}^b{\Sigma_\alpha}^{\rho}+A^{a}{}_{\nu b}{\Lambda_{\mu}}^b{\Sigma_a}^{\rho}]\notag\\
 \label{PP}
\end{align}
Using ${\Sigma_a}^{\rho}=h^{\rho\sigma}{\Lambda_{\sigma}}^a$ (which follows from (\ref{spm})) and (\ref{tem}), (\ref{tm2}), the above expression will take the form as,
\begin{align}
\Gamma_{\nu\mu}^{\rho}&=\tau^{\rho}\partial_{(\mu}\tau_{\nu)}+
\frac{1}{2}h^{\rho\sigma}[\partial_{\mu} h_{\sigma\nu}-{\Lambda_{\nu}{}^a}\partial_{\mu}{\Lambda_{\sigma}}^a]+\frac{1}{2}h^{\rho\sigma}[
\partial_{\nu} h_{\sigma\mu}-{\Lambda_{\mu}{}^a}\partial_{\nu}{\Lambda_{\sigma}}^a]\notag\\& +A^{a}{}_{0\mu}{\Lambda_{\nu}}^{0}{\Sigma_\alpha}^{\rho}+A^{a}{}_{0\nu}{\Lambda_{\mu}}^0{\Sigma_a}^{\rho}+A^{a}{}_{\mu b}{\Lambda_{\nu}}^b{\Sigma_\alpha}^{\rho}+A^{a}{}_{\nu b}{\Lambda_{\mu}}^b{\Sigma_a}^{\rho}
\label{comid}
\end{align}
To get the desired expression for the connection we need a little more algebra. Exploiting the symmetricity of $\Gamma_{\nu\mu}^{\rho}$ we can write,
\begin{equation}
\frac{1}{2}h^{\rho\sigma}[-{\Lambda_{\nu}{}^a}\partial_{\mu}{\Lambda_{\sigma}}^a
-{\Lambda_{\mu}{}^a}\partial_{\nu}{\Lambda_{\sigma}}^a]=-\partial_{\sigma}h_{\mu\nu}-A^{a}{}_{\mu b}{\Lambda_{\nu}}^b{\Sigma_a}^{\rho}-A^{a}{}_{\nu b}{\Lambda_{\mu}}^b{\Sigma_a}^{\rho}
\label{conec}
\end{equation}
Finally, Using (\ref{conec}), we obtain the cherished form of the connection from (\ref{comid}) that resembles the structure (\ref{covariantconnection}),
\begin{align}
{\Gamma^\rho}_{\nu\mu} & = \tau^{\rho}\partial_{(\mu}\tau_{\nu)} +
\frac{1}{2}h^{\rho\sigma} \Bigl(\partial_{\mu}h_{\sigma\nu}+\partial_{\nu}h_{\sigma\mu} - \partial_{\sigma}h_{\mu\nu}\Bigr)+ h^{\rho\lambda}K_{\lambda(\mu}\tau_{\nu)}
\end{align}
where the two form K is defined as,
\begin{align}
 h^{\rho\lambda}K_{\lambda(\mu}\tau_{\nu)} &=\frac{1}{2}h^{\rho\lambda}[K_{\lambda\mu}\tau_{\nu}+K_{\lambda\nu}\tau_{\mu}]\notag\\
 &=\frac{1}{2}[A^{a}{}_{0\mu}{\Lambda_{\nu}}^{0}{\Sigma_\alpha}^{\rho}+A^{a}{}_{0\nu}{\Lambda_{\mu}}^0{\Sigma_a}^{\rho}]
\end{align}
\section{Conclusion} 
Newtonian gravity was cast as a geometric theory on 4-dimensional spacetime by Elie Cartan \cite{Cartan}, and was subsequently developed by many stalwarts \cite{TrautA} - \cite{EHL}. Interest in the problem has been rekindled due to the possible applications of nonrelativistic diffeomorphism in NR superparticle theory \cite{BGJKM}, fractional Quantum Hall effect \cite{SW} and also in holographic theories \cite{CMH}. Furthermore, a novel formalism for the emergence of spatial diffeomorphism was developed by localising the galilean symmetry of nonrelativistic field theories \cite{BMM}.

    In this paper we have presented a complete account of the localisation of the global galilean invariance of nonrelativistic field theories leading to a dynamical realization of the Newton-Cartan geometry. The fields introduced during the localisation of the galilean symmetry were used to build the structures of this geometry. It is a new formulation which differs fundamentally from the earlier works. For instance, gauging of galilean symmetry of a non relativistic particle model was performed earlier in \cite{PLP} but the gauge fields were not utilised to link with Newton-Cartan gravity. Rather the gravitational dynamics of the gauge fields was obtained from the contraction of a relativistic particle model. In another approach, gauging of the galilean algebra was performed from a contraction of the Poincare algebra \cite{BP}. 

 A general Lagrangian field theory invariant under the galilean group of transformations has been considered. The parameters of this group of transformations are constants. Localisation of these transformations means to make the parameters functions of space and time. Naturally the original theory is no longer invariant under these local galilean transformations. The invariance is brought back in two steps. First, locally covariant derivatives were obtained that transformed under the local transformations in the same way as the partial derivatives under global transformations. Secondly, the integration measure was required to be suitably changed.
 
  So far our attention was confined entirely within nonrelativistic field theories in Euclidean space and universal time. For vanishing time translation parameter our theory, invariant under local rotation and  boosts, was demonstrated to be equivalent to field theory in curved space. This means that a general prescription was obtained to formulate a nonrelativistic diffeomorphism invariant theory. As already mentioned such theories have recently been employed for analysing fractional quantum Hall effect. 

The geometrical analogy was then pushed forward in a big way. Introducing a 4-dimensional manifold we were able to identify the vierbein fields from the set of fields introduced during gauging. Using only the vierbein postulate we were able to endow the  4-dimensional manifold with structures that makes it equivalent to the Newton-Cartan spacetime. It was indeed gratifying to observe how the transformation rules obtained during the localisation procedure provided the correct geometrical transformations to the various objects of the Newton-Cartan spacetime.

\end{document}